\documentclass{webofc}

\usepackage[varg]{txfonts}   
\usepackage{bm}
\usepackage{amsmath,amssymb,mathrsfs,amsfonts,slashed}
\usepackage{hyperref}
\usepackage{url}
\usepackage{wrapfig}

\newcommand{\xscapegim}{\textsc{x-scape-gim}}

\hypersetup{colorlinks=true,citecolor=blue,urlcolor=blue,linkcolor=blue}
\newcommand{\p}{\bm{p}}
\newcommand{\x}{\bm{x}}
\begin{document}
\title{Jets: Hard-soft correlation - Theory overview}

\author{\firstname{Ismail} \lastname{Soudi}\fnsep\thanks{\email{ismail.i.soudi@jyu.fi}}
}

\institute{
    University of Jyväskylä, Department of Physics, P.O. Box 35, FI-40014 University of Jyväskylä, Finland.
    \and
    Helsinki Institute of Physics, P.O. Box 64, FI-00014 University of Helsinki, Finland.
}

\abstract{Overview of recent theoretical advances presented at Hard Probes 2024 in the study of hard-soft correlation in heavy ion collisions and small systems.}
\maketitle
\section{Introduction}\label{sec:intro}
The measurement of QCD jets have been one of the main driving force in the study of high-energy nuclear collisions.
In $p$-$p$ collisions, jets are used to understand fundamental properties of proton structure and to test perturbative QCD\@.
In heavy-ion collisions, the observation of jet quenching is a strong signature of Quark-Gluon-Plasma (QGP) formation.
By studying this modification of jets, we hope to probe the properties of the QGP\@.

Due to the large momentum transfer, hard processes occur during short times scales, producing hard partons at early stages of the collision.
These hard partons originate with high virtuality which lead to a vacuum-like fragmentation lowering their virtuality.
While these fragments are directly detected as collimated jets in vacuum, in the presence of a medium, they must pass through the plasma before reaching the detector.
This interaction with the medium leads to energy loss and modification of the jet structure.

Since the formation of vacuum-like emissions is fast, they are expected to be factorized from medium interactions~\cite{Caucal:2018dla}.
However, as the vacuum-like shower evolves, the resolution scale of the medium becomes more relevant.
Several works have investigated how the medium can start to affect the vacuum-like shower at later stages of the shower evolution~\cite{Caucal:2018dla,Kumar:2019uvu,Mehtar-Tani:2024jtd}.
It was shown that the precise moment at which medium interactions become significant is crucial for accurately describing both hard parton suppression and their azimuthal asymmetries~\cite{Andres:2019eus}.

Once the parton virtuality scale becomes comparable to medium scales, the medium starts to resolve individual emissions.
This leads to scattering with the medium which broaden the jet core.
Moreover, due to multiple soft scattering with the medium, the hard partons are driven slightly off-shell which leads to medium-induced radiation.
This radiation is the dominant mechanism for energy loss in the medium.
In the literature, two primary approaches have been employed to describe jet-medium interactions:
A strong coupling approach based on the AdS/CFT correspondence~\cite{Liu:2006ug}, and a weak coupling approach based on perturbative QCD (pQCD) which is the focus of this manuscript.

The QGP undergoes a complex evolution with different stages throughout the collision.
After reaching local thermal equilibrium, the QGP has been successfully described by hydrodynamical simulations.
The energy and momentum deposited by the hard partons in the medium leads to perturbations on top of the hydrodynamical background.

In small systems such as $p$-$A$ and $p$-$p$ collisions, recent measurements of $v_2$, the elliptic flow coefficient have observed signatures of collective flow reminiscent of heavy ion collisions~\cite{ATLAS:2017hap,ALICE:2014dwt,CMS:2015yux}.
However, there has been no indication of jet quenching in these systems.
Due to limited energy budget in small systems, understanding the correlations between the hard and soft sector is crucial to understanding these recent measurements.

\section{Jets in heavy ion collisions}\label{sec:large}
Jet-medium interactions can be described using an effective kinetic theory, where the partons are treated as on-shell quasi-particles.
The evolution of the phase-space distribution is described by the Boltzmann equation~\cite{Arnold:2000dr,Arnold:2001ba,Arnold:2003rq}
\begin{align}
    \left(\partial_t + \frac{\p}{|\p|} \bm{\nabla}_x\right) f_a(\p,\x,t) 
    = C_a^{2\leftrightarrow2}[\{f_i\}] + C_a^{1\leftrightarrow2}[\{f_i\}]\;,
\end{align}
At leading order of pQCD, two main processes contribute to the energy loss of hard partons: number conserving elastic scatterings $C_a^{2\leftrightarrow2}$ and radiation induced by the medium which can be resummed into an effective collision term $C_a^{1\leftrightarrow2}$~\cite{Arnold:2000dr,Arnold:2001ba,Arnold:2003rq}.

In the weak coupling approach, the process responsible for energy loss are the same that drives the initial highly energetic plasma to local equilibrium~\cite{Baier:2000sb}.
Accordingly, the evolution of the medium component $f^{\rm med}$ can itself be described by the same Boltzmann equation.
The phase-space distribution can be separated into medium and jet components
\begin{align}
    f(\p,t) = f^{\rm med}(\p, t) + \delta f^{\rm jet}(\p, t) \,.
\end{align}
Since the jet is dilute compared to the dense QGP, the jet distribution is treated as a perturbation on top of the medium distribution.
The Boltzmann equation can be linearized around the medium, ignoring any subleading jet-jet interactions.

Energy loss has typically been studied using MonteCarlo stochastic approaches~\cite{Zapp:2012ak,Zapp:2013vla,Wang:2013cia,He:2015pra,Yazdi:2022bru,Shi:2022rja}.
In order to describe medium response, one then needs to couple the jet evolution to a hydrodynamical simulations.
By introducing a source term in the hydrodynamic simulation, the energy-momentum lost by the jet is deposited in the medium~\cite{Stoecker:2004qu,Casalderrey-Solana:2004fdk,Chaudhuri:2005vc,Betz:2008ka,Tachibana:2020mtb}.
Another approach is to directly solve the linearized Boltzmann equation.
This allows the treatment of the full energy cascade from the high energy partons $\sim E$ to medium scales $\sim T$ using the same framework~\cite{Schlichting:2020lef,Mehtar-Tani:2022zwf,Zhou:2024ysb}.

\paragraph{Static background:}\label{sec:static}
\begin{figure}
    \centering
    \includegraphics[width=0.45\textwidth]{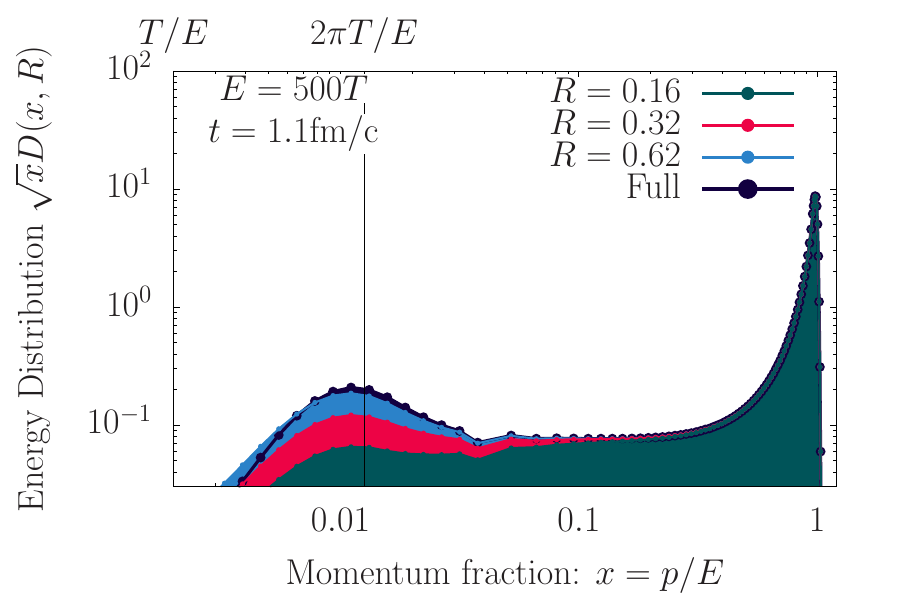}%
    \includegraphics[width=0.45\textwidth]{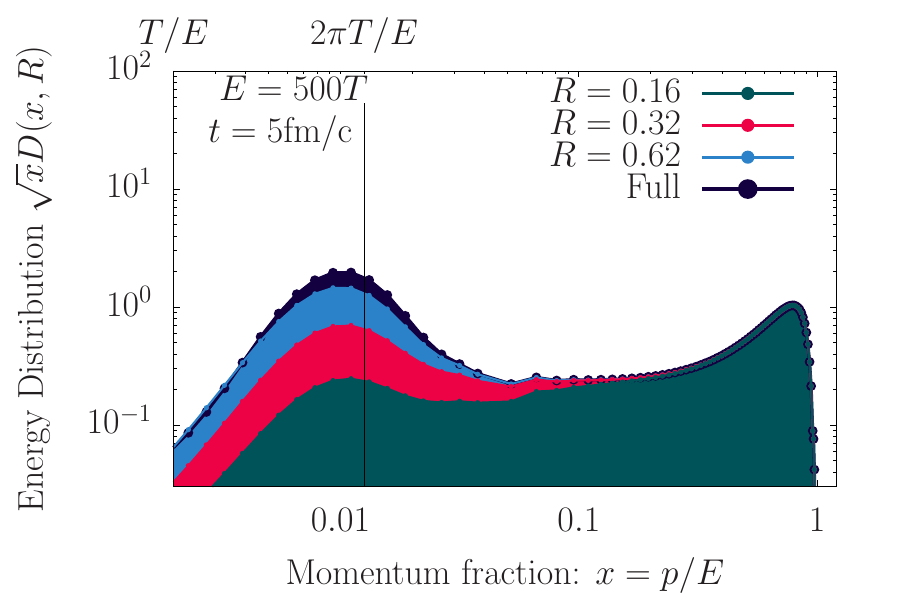}
    \caption{Evolution energy distribution with a decomposition into different cone sizes $R=0.16, 0.32, 0.62$ and $\pi$.
        Figure from Ref.~\cite{Mehtar-Tani:2024jtd}.
    }\label{fig:Static}
\end{figure}

The evolution of the jet energy distribution $D(x) = \int dp \delta(p-xE) p^3 \delta f(p)$, with initial energy $E$ in static equilibrium background is depicted in Fig.~\ref{fig:Static}. 
The collinear emissions dominate the energy loss, leading to efficient transport of energy from the hard scales $\sim E$ to the medium scales $\sim T$ without deposition in the intermediate scales.
The energy cascade can be described using Kolmogorov-Zakharov~\cite{Zakharov:1992} wave turbulence characterized by a stationary solution in the intermediate scales $D(x) \simeq 1/\sqrt{x}$~\cite{Blaizot:2012fh,Blaizot:2014bha}.
Elastic scattering are responsible for the redistribution of the energy deposited in the medium out to large angle.

\paragraph{Dynamic background:}
Recent works have studied the equilibration of jets in a dynamic background where the QGP distribution itself is equilibrating~\cite{Zhou:2024ysb}.
Starting with highly occupied plasma, the QGP undergoes a non-equilibrium evolution approaching a viscous hydrodynamics description.
The pressure anisotropy of the jet distribution $A = \frac{\delta P_T - \delta P_L}{\delta e /3}$ is used to characterize the hydrodynamisation.
Following jets with different initial momenta with angles $\theta$ with respect to the initial QGP anisotropy.
They find that the medium expansion leads to a rapid decrease of the longitudinal pressure $P_L$.
The different jets approach the same hydrodynamical limit already at early times.

If the medium is large, all the jet fragments will eventually thermalize with the medium at asymptotic times.
The only signal of the hard parton will be a change of the thermodynamic variables (e.g.~temperature, flow velocity,\ldots).
For the case of a static background, the asymptotic energy distribution is obtained by matching the leading order Taylor expansion in the thermodynamic variables of the equilibrium distribution~\cite{Mehtar-Tani:2024jtd}
\begin{align}
    D^{(\rm eq)}_{a/jet}(x,\theta)
    \simeq& \left(1+3\cos\theta\right) \partial_T n_a(xE)\;.
\end{align}
Here $\theta$ is the azimuthal angle of the initial jet momentum and $n_a$ is the equilibrium distribution of the medium.
This leads to a boosted energy distribution along the jet direction.

\section{Jets in small systems}\label{sec:small}
In heavy ion collisions, the energy carried by the soft sector is much larger than the hard partons' energy.
This allows the treatment of hard partons as a small perturbation on top of the medium.
However, in small systems, the presence of a hard parton with energy $\gtrsim 100$~GeV significantly alters the remaining energy of the soft bulk.

There have been no observation of suppression in measurements of the $R_{AA}$ for minimum bias $p$-$Pb$~\cite{CMS:2016svx,ALICE:2018vuu,ATLAS:2022kqu}.
However, when selecting for centrality, measurements show suppression for central events and enhancement for more peripheral events~\cite{ATLAS:2014cpa}.
This hints at a possible interplay between the hard and soft particle production.

\paragraph{Hard-soft correlation: x-scape:}

The \xscapegim{} described in~\cite{JETSCAPE:2024dgu} employs a multi-stage approach that imposes event-by-event energy-momentum conservation.
The energy-momentum consumed in hard processes is subtracted from that available for bulk medium production.
The hadron and jet spectra in minimum-bias $p$-$p$ and $p$-$Pb$ collisions reproduce experimental data across the full transverse momentum range at top LHC energies.

Energy loss is found to be negligible in $p$-$Pb$ collisions, aligning with experimental findings.
Correlations between the hard and soft particle production are characterized by measuring the mean transverse energy $\langle E_T \rangle$ as a function of leading jet $p_T$.
At mid-rapidity the mean transverse energy $\langle E_T \rangle$ increases monotonically as a function of leading jet $p_T$.
For forward rapidities, it was shown that the $\langle E_T \rangle$ is peaked at $p_T \simeq 100$~GeV.
While it increases with $p_T$ at low $p_T$, the competition between the hard and soft sector leads to a decrease of $\langle E_T \rangle$ as one selects events with larger jet $p_T$.
These results qualitatively explain the correlations between the centrality selection and hard process triggers observed by ALICE and ATLAS measurements.

\paragraph{TMD-PDFs:}
\begin{wrapfigure}{L}{0.45\textwidth}
    \includegraphics[width=0.45\textwidth]{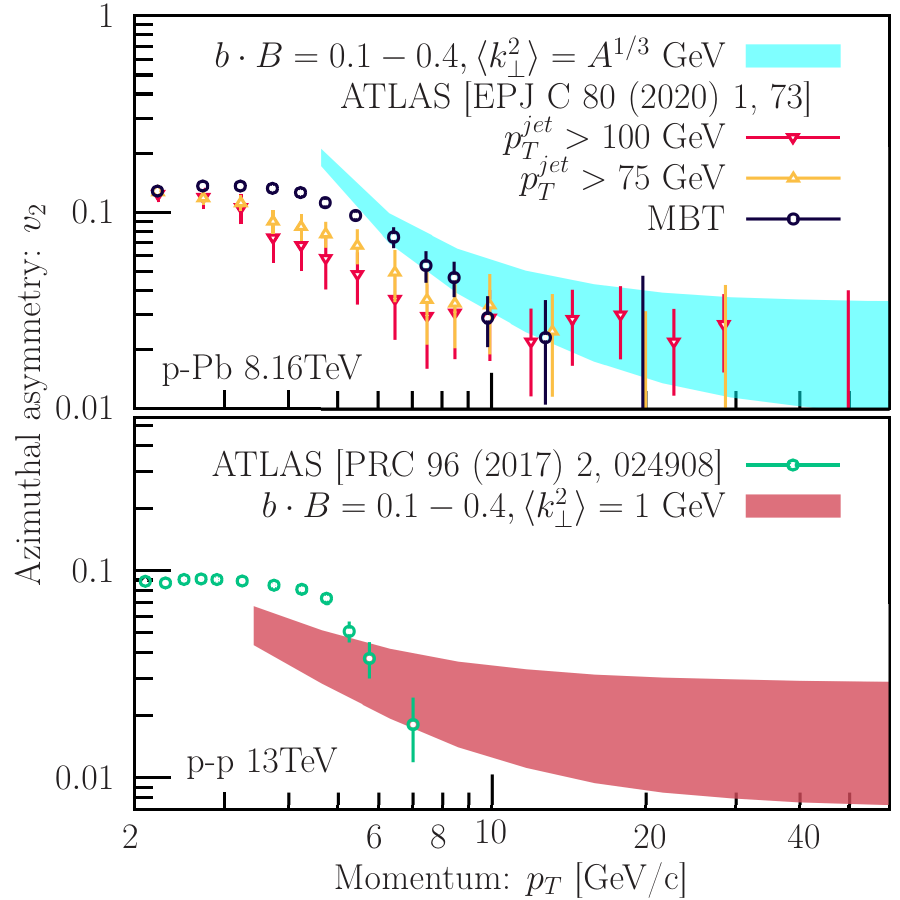}
    \caption{Azimuthal anisotropies obtained using TMD distribution figure from Ref.~\cite{Soudi:2023epi}.}\label{fig:v2}
\end{wrapfigure}

Recent work have shown that azimuthal anisotropies observed in $p$-$p$ and $p$-$Pb$ at high-$p_T$ can be attributed to initial state transverse momentum correlations~\cite{Soudi:2023epi,Soudi:2024slz}.
Consider a hard scattering process with total transverse momentum $q_T$, due to momentum conservation, it must be balanced by the total transverse momentum of the remaining partons contributing to soft particle production.
In the final state, the net transverse momentum of the produced dijet is balanced by the net transverse momentum of the bulk system.

One can then compute the azimuthal anisotropies as the Fourier decomposition of the momentum in-balance of dijet production.
Azimuthal correlations generated from transverse momentum dependent parton distribution functions (TMDPDFs) and fragmentation functions (TMDFFs) leads to sizeable $v_2$ at high-$p_T$ as shown in Fig.~\ref{fig:v2}.

\paragraph{Bjorken-$x$:}
In order to compute hadron suppression in heavy ion collisions, experiments typically use a Glauber model to estimate the number of binary collisions.
While in $Pb$-$Pb$ collisions there is a clear correlation between the multiplicity and number of particles, in small systems large relative statistical fluctuations per nucleon-nucleon collision lead to selection biases~\cite{ALICE:2014xsp}.
In order to mitigate these biases PHENIX has made measurements comparing the yield of neutral pion and direct photon production in central $d$-$Au$ collisions~\cite{PHENIX:2023dxl}.
The photon yield is used to correct for any centrality bias effects.
Using the initial-state color fluctuation model~\cite{Alvioli:2013vk}, it was shown that the observed suppression can be explained by initial Bjorken-$x$ correlations without the need of final-state interactions~\cite{Perepelitsa:2024eik}.

\section{Summary}
Jets constitute a powerful tool to study QCD matter in relativistic hadron colliders.
Hard partons energy loss is dominated by medium-induced radiation.
Once the inverse energy cascade transports the energy to medium scales, these perturbations are thermalized with the medium.
Understanding thermalization of these soft modes is crucial for understanding medium response to hard probes.

In small systems, correlations between the hard and soft particle production become more important.
Energy-momentum conservation in \xscapegim{} leads to interplay between the hard and soft particle production.
While azimuthal correlations in the initial state can explain the observed final state anisotropies at high-$p_T$.
Moreover, the color fluctuation model can explain the observed suppression in $d$-$Au$ collisions without the need of final-state interactions.

Since heavy ion collisions show clear evidence of jet quenching and small systems do not.
Future light-ion collisions such as $O$-$O$ collisions at the LHC will be important to understand the transition between these two regimes.

\section*{Acknowledgements}
The author is funded as a part of the European Research Council project ERC-2018-ADG-835105 YoctoLHC, and as a part of the Center of Excellence in Quark Matter of the Academy of Finland (Projects No. 346325 and 364192).
%
\bibliography{ref} 
\end{document}